\documentclass[11pt,a4paper]{article}
\usepackage{jheppub}
\usepackage{amsmath,amssymb,amsthm, epsf,yfonts}
\usepackage{graphicx}  
\usepackage{epsfig,subfigure,placeins,float} 
\usepackage{bm}        
\usepackage{verbatim}   
\usepackage{color}
\usepackage{hyperref}
\hypersetup{colorlinks,linkcolor={red},citecolor={blue},urlcolor={blue}}  
\usepackage{slashed}
\usepackage{amssymb}
\usepackage{bbold}
\usepackage{footnote}
\usepackage{float}
\usepackage[]{natbib}
\RequirePackage{filecontents}
\newcommand{\bca}{\begin{cases}}
\newcommand{\eca}{\end{cases}}

\title{Probing the Holographic Fermi Arc with scalar field: Numerical and analytical study}
\author[1]{Sayan Chakrabarti, \note{Corresponding author.}}
\author{Debaprasad Maity}%
\author{and Wadbor Wahlang}
\affiliation{Department of Physics,\\
   Indian Institute of Technology, Guwahati\\
Guwahati-781039, India}
\emailAdd{sayan.chakrabarti@iitg.ac.in}
\emailAdd{debu@iitg.ac.in}
\emailAdd{wadbor@iitg.ac.in}
\abstract{Fermi arcs are disconnected contour of Fermi surface, which can be observed in pseudo-gap phase 
of high temperature superconductors. Aiming to understand this pseudo-gap phenomena, we study a holographic Fermionic system coupled with a massive scalar field in an AdS black hole background. Depending on the boundary condition on the scalar field mode, we discuss two possible scenarios. When the scalar condenses below a critical temperature $T_c$, Fermi surface undergoes a transition from normal phase to pseudo-gap phase. Hence $T_c$ can be the reminiscent of well known cross over temperature $T^*$ in cuprate superconductor, below which pseudo-gap appears at constant doping. In the second scenario, the bulk scalar develops a non-normalizable profile at arbitrary temperature for non-zero source at the boundary. Therefore, we can tune the Fermi spectrum by tuning a dual source at the boundary. The dual source for this case can be the reminiscent of hole doping in the real cuprate superconductor. For both the cases we have studied Fermi spectrum and observed anisotropic gap in the spectral function depending on the model parameter and studied the properties of Fermi arcs across different phases. }
\arxivnumber{1902.08826 }

\begin{document}
\maketitle
\flushbottom

\section{Introduction}

Over the last decade, the holographic principle has evolved as a tool to describe the strongly coupled systems which otherwise were hard to study in a general perturbative approach. Holographic principle essentially relates the physics behind strongly interacting quantum systems to that of weakly interacting gravitational theory with one higher dimension. It has been first used in the context of $\mathcal{N}=4$ SYM theories \cite{Witten:1998qj,Maldacena:1997re}. However, the fact that the holographic principle could be of great help to model the real-world systems as well was realised afterwards.  Accordingly, applications towards strongly coupled systems in quantum chromodynamics were studied in great detail. On the other hand, the holographic principle is also being used as a framework to understand non-linear hydrodynamics \cite{Bhattacharyya:2008jc}, Fermi liquid behavior \cite{Faulkner:2009wj,Cubrovic:2009ye}, transport phenomena \cite{Herzog:2007ij}, high temperature superconductors \cite{Hartnoll:2008vx,Hartnoll:2008kx,Horowitz:2010gk} to name a few, thereby exploring condensed matter systems with the help of classical gravity theories. Interested readers are referred to the articles \cite{Hartnoll:2009sz,Herzog:2009xv,McGreevy:2009xe,Zaanen:2015oix,Hartnoll:2016apf} for detailed reviews in this regard. It is therefore safe to say that the holographic principle has become an extremely powerful method to study strongly-correlated systems, be it in high energy physics or in condensed matter theory. One of the important aspects of condensed matter systems is to understand Fermionic system in strong coupling regime. A quantity of interest in strongly coupled condensed matter systems, which can be
computed using holography, is the spectral function of the Fermion, which in turn is proportional to the imaginary part of the Fermionic retarded two-point correlation function. A lot of works were done in this regard in the literature \cite{Liu:2009dm,Faulkner:2009wj,Lee:2008xf}, where the Dirac equation for a charged probe Fermion propagating in a gravitational background is analysed. Many interesting and emergent phenomena seems to emerge out of these retarded two-point function of the dual Fermionic operator in the boundary theory.

The discovery of superconductivity, on the other hand,  in the LaBaCuO ceramics at $30$ K by Bednorz and M\"{u}ller \cite{Bednorz:1986tc} in 1986 has opened the era of high-$T_c$ superconductivity. Prior to this, the phenomena of superconductivity had been confined to very low temperatures. This unexpected result prompted intense activity in the field of ceramic oxides, both in the experimental as well as theoretical front. These ceramic oxides have a superconducting phase with an order parameter having $d$-wave symmetry. This phase exists for the hole-doped material over $5\%$ range of doping. However the material behaves like an antiferromagetic Mott insulator for very low dopings. These two states are connected by an unconventional phase known as the pseudo-gap. Angle Resolved Photoemission Spectroscopy (ARPES) \cite{Damascelli:2003bi} has found the presence of a truncated Fermi surface in the pseudo-gap phase and this truncated Fermi surface in the momentum space is termed as a Fermi Arc \cite{1998Natur.392..157N,2006PhRvB..74v4510Y,2011PhRvL.106l7005K,Cremonini:2018xgj,Seo:2018hrc}. This phenomenon remains a mystery, mostly because of the unconventional electronic properties of the normal state of the superconductor. As is well known, Landau's theory of metals predicts continuous closed Fermi surfaces and does not explain the observed truncated surfaces in momentum space when holes are doped into the copper-oxide plane.  The arc is intermediate between the $d$-wave node of the superconductor and the complete Fermi surface of the normal state of the superconductor. On top of that, the arc appears to be formed by a closing of the energy gap of the superconducting state when temperature is increased above $T_c$.  

The deviation from standard Landau's theory of metal is generally taken into account in two different types of explanations \cite{Vanacore:2015poa}. It is shown that some type of order can set in \cite{PhysRevB.76.174501} to give rise to a Fermi pocket with momentum dependent spectral intensity. This intensity can be extremely small (almost close to zero) for some range of momenta, giving rise to the discontinuity of the Fermi surface and thereby generating an arc-like structure in the momentum space. Secondly, the phenomenon was tried to be explained keeping the inherent strong coupling nature of the problem in mind, in which zeros of the single-particle electronic Green function, caused essentially by a divergent self-energy \cite{PhysRevB.73.174501,PhysRevB.86.115118} are the reasons for the vanishing of the surface. It is important to note that Fermi arcs have been obtained phenomenologically \cite{PhysRevB.73.174501} and numerically \cite{PhysRevB.74.125110} from the point of view of condensed matter physics, as well as there have been holographic description of it too \cite{Vegh:2010fc,Benini:2010qc}. In these holographic descriptions, the arcs were obtained by condensing the Fermions anisotropically into $p$-wave or $d$-wave superconducting states, but these does not describe the cuprates in the pseudo-gap phase. 

It is well known that one can implement the holographic prescription in different ways for finite density Fermionic system. In this paper, we follow the program described in \cite{Vanacore:2015poa}, where, the action for a bulk gravitational system is supplemented with Fermionic fields which act as source to a Fermionic operator at the boundary of AdS.
It has been shown \cite{Faulkner:2009wj} already that a simple canonical Fermion field in the Reissner-Nordstr\"{o}m ${\rm{AdS_4}}$ black hole background can give rise to both Fermi liquid and non-Fermi liquid behavior at the boundary.
Although, this kind of constructions \cite{Faulkner:2009wj,Cubrovic:2009ye} can give rise to gapped spectra depending upon the bulk Fermion mass, it is very difficult to produce pseudo-gap spectrum without invoking any new coupling. One possible mechanism of obtaining pseudo-gap proposed by Vanacore et. al. \cite{Vanacore:2015poa} is by considering a non-minimal Fermion and gauge field interaction. The purpose of this paper is to generalize and most importantly dynamically generate the aforementioned Fermion-gauge interaction by introducing another neutral scalar field and study its effect on the Fermi arc. We will consider two different scenarios. Firstly, scalar field in the bulk can act as a neutral order parameter field giving rise to transition from normal phase to pseudo-gap phase. The phase transition temperature of this scalar field may be identified with the well known crossover temperature $T^*$ in the high-$T_c$ superconducting phase diagram, below which effects of electronic pairing correlations become significant\cite{PhysRevB.54.R3756,RANDERIA19981754}.
Therefore, in the holographic model this scalar field under consideration could be related to the pairing phase fluctuation from the dual field theory point of view. However, to confirm such claim we need to have detailed study on this issue. In the second scenario corresponding to the scalar field, we assume the existence of a dual tuning scalar operator which controls the pseudo-gap phase. From our discussion, it appears that the same neutral scalar field is responsible for both the aforementioned mechanisms. However, strictly speaking this is not the case. Even though the theory of scalar field looks same for both the mechanisms, in principle they are different. The difference can be easily understood if we consider the mass of the scalar field to be above $AdS_2$  Breitenlohner-Freedman (BF) bound $m_{\Phi} >-3/2$. The scalar field satisfying this condition will not condensate at any temperature. Therefore, the first mechanism will be not realisable for this system at all. Detail field theoretical understanding of these two scenarios could be an important topic of further research.

We organise our paper as follows: in section II and III we will present a very brief discussion of the background geometry and the scalar field solution followed by section IV where we discuss the Fermionic Lagrangian. We present our numerical results, a discussion on the energy gap in the spectral function and analytical discussion of the spectral function in sections V, VI and VII respectively. Finally, in section VIII we conclude the paper with a brief discussion and future directions. 

\section{Review of Background Geometry and Action}

  Taking the simplest action coupled to gravity in AdS$_4$ with real massive scalar and gauge field
\begin{align}\label{action}
 \mathcal{S}=\frac{1}{2\kappa^2}\int d^4x\sqrt{-g}\left[\mathcal{R}+\frac{6}{L^2}-\frac{1}{4}F^2+\frac{1}{\lambda}\left(-\frac{1}{2}g^{\mu\nu}\nabla_{\mu}\Phi\nabla_{\nu}\Phi-V(\Phi)\right)\right]
\end{align}
with $L$ the AdS curvature radius and $F=dA$. Here $\lambda$ is a coupling constant and the potential $V(\Phi)$ is given by 
 \begin{align*} 
 V(\Phi)=\frac{1}{4L^2}\left(\Phi^2+m^2_{\Phi}L^2 \right)^2-\frac{m^4_{\Phi}L^2}{4}.
 \end{align*}
 The main motivation of choosing the above form of the potential is to have a non trivial scalar field solution, where $m_{\Phi}$ is scalar field mass. In this paper we will neglect the effects of backreaction from the scalar field by taking $\lambda$ to be large but include the effects of the gauge field leading to our electrically charged AdS black hole. For a detailed discussion the reader is referred to \cite{Iqbal:2010eh}. 
 The equations of motion obtained from action (\ref{action}) are
\begin{align}
&R_{\mu\nu}-\frac{1}{2}g_{\mu\nu}R -\frac{3g_{\mu\nu}}{L^2} =\frac{1}{2}F_{\mu\lambda}F^{\lambda}_{\nu}-\frac{1}{8} g_{\mu\nu}F^2+\nabla_{\mu}\Phi\nabla_{\nu}\Phi+g_{\mu\nu}\left(-\frac{1}{2}\nabla_{\rho}\Phi\nabla^{\rho}\Phi-V(\Phi)\right), \nonumber \\
& \frac{1}{\sqrt{-g}}\nabla_{\mu}\big(\sqrt{-g}g^{\mu\nu}\nabla_{\nu}\Phi\big)-\frac{1}{L^2}(\Phi^2+m_{\Phi}^2L^2)\,\Phi=0.
\end{align}
On the other hand, it is well known that black holes are the simplest objects in general relativity. Assumption of rotational symmetry leads to the fact that the geometry of the black hole is fully specified by its mass and charge, independent of other details of the system. Systems which are of relevance in condensed matter physics are mostly finite density systems, with temperature much smaller than the chemical potential. These type of systems in the  gravity side are described by black holes having charges. We will therefore, choose a familiar Reissner Nordstr\"{o}m (RN) $AdS_4$ black hole as our background geometry with $\Phi=0$. The general $AdS_4$ metric  is given by the following line element: 
\begin{equation}\label{metric}
 \frac{ds^2}{L^2} = -g_{tt}dt^2 + g_{rr}dr^2+g_{xx}dx^2+g_{yy}dy^2\,.
\end{equation}
For RN-AdS$_4$, the metric coefficients are given by
\begin{align*}
 g_{tt}=r^2f(r)\,\,,\,g_{rr}=r^{-2}f^{-1}(r)\,\,\text{and}\,\, g_{xx}=g_{yy}=r^2,
\end{align*}
where, after rescaling, the horizon is at $r=1$ and all coordinates are dimensionless. The metric function and the $U(1)$ gauge field $A_t$ are given by:
\begin{equation*}
 f=1 + \frac{3\gamma}{r^4}-\frac{1+3\gamma}{r^3}, \hspace{0.1cm} A_t=\mu(1-\frac{1}{r})dt.
\end{equation*}
Here, we expressed the chemical potential \cite{Iqbal:2010eh}, as $\mu \equiv\sqrt{3}\gamma^{\frac{1}{2}}$ and the black hole temperature as $T=\frac{3}{4\pi}(1-\gamma)$. The parameter $\gamma$ ranges from 0 to 1 and it controls the temperature and chemical potential of the system. For $\gamma=1$, we have the extremal black hole with $T=0$, and for $\gamma=0$, we have finite temperature system with zero chemical potential ($\mu=0$) and hence zero charge density on the dual field theory.

\section{Scalar Field Solution}
As we have already described, the scalar field in our model plays very distinct role in controlling the properties of Fermi arc. Properties of solution of such a field in the RN black hole background is well studied. For completeness let us discuss about the important behaviour of it.
 The radial equation of motion for $\Phi(r)$ is given by

\begin{figure*}[htbp]
 \centering
\includegraphics[scale=0.5]{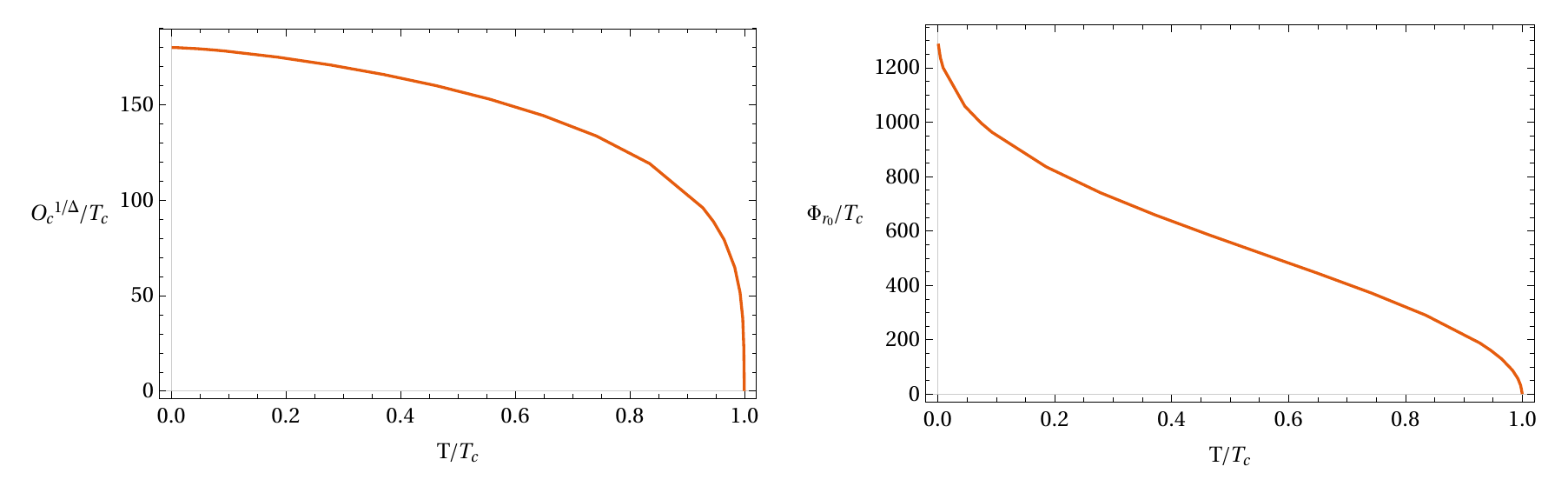}
 \caption{Plot of Condensate $O_c$\,(left) and horizon value of $\Phi$\,(right) vs temperature ($T/T_c$) with $O_s=0$. Here $m_{\Phi}^2=-21/10$ and $T_c\approx$0.001078. We also found that $T_c$ decreases when $m_{\Phi}^2$ approach the BF bound.}
\label{fig1}
 \end{figure*}
 \begin{figure}[htbp]
 \centering
\includegraphics[scale=0.5]{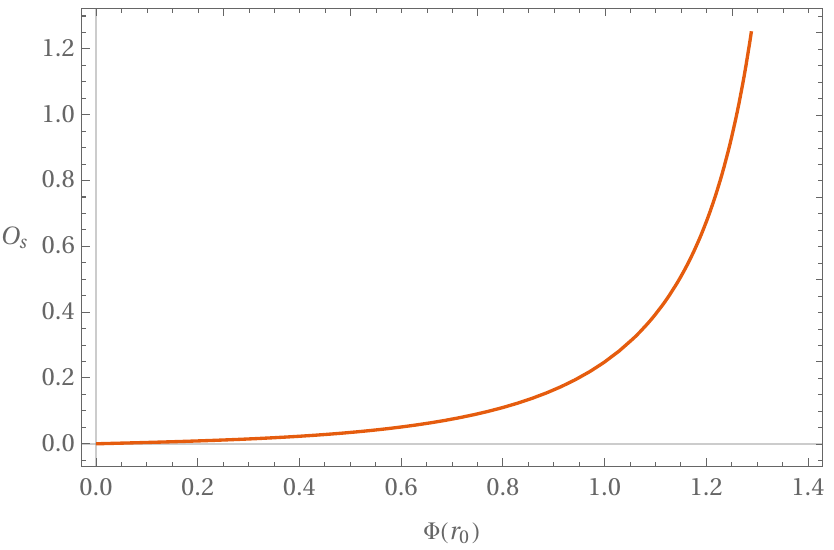}\hspace{0.4cm}
 \caption{Plot of source $O_s$ vshorizon value of $\Phi(r_0)$ for $T>T_c$.}
\label{fig2}
 \end{figure}
\begin{align}\label{sceqn}
\Phi''(r)+\left(\frac{f'}{f}+\frac{4}{r}\right)\Phi'(r)-\frac{\left(\Phi(r)^2+m_{\Phi}^2L^2\right)}{r^2f}\Phi(r)=0.
\end{align}
From this, the asymptotic behaviour of $\Phi$ can be written as 
\begin{equation}\label{asymtoticphi}
\lim_{r\rightarrow \infty}\Phi(r)=\frac{O_s}{r^{3-\Delta}}+\frac{O_c}{r^{\Delta}},
\end{equation}
where $
\Delta={3}/{2}+\sqrt{{9}/{4}+m_{\Phi}^2L^2}$ is identified with the scaling dimension of the dual field theory operator.
In the above expression, we set $L=1$. As has already been emphasised, one of our goal is to study the evolution of Fermionic spectral function across the phase transition, we therefore chose the mass of the scalar field within BF bound $-\frac{9}{4}<m_{\Phi}^2<-\frac{3}{2}$. This essentially violates the $AdS_2$ BF bound near the horizon at zero temperature.  
For normal quantisation we interpret $O_s$ as the source and $O_c$ as response.
From equation (\ref{asymtoticphi}) we can write 
\begin{align}
\lim_{r\rightarrow \infty}\Phi(r)&=O_s\, r^{\Delta -3}+O_c\, r^{-\Delta }\,,\nonumber\\
\lim_{r\rightarrow \infty}r\Phi'(r)&=O_s (\Delta -3)\, r^{\Delta -3}-O_c\, \Delta\,  r^{-\Delta }.
\end{align}
Solving for $O_s$ \& $O_c$ we get
\begin{align*}
 O_s&=\lim_{r\rightarrow \infty}\left[\frac{r^{3-\Delta } \left(\Delta\,\Phi (r)+r\,\Phi '(r)\right)}{2 \Delta -3}\right],\\
 O_c&=\lim_{r\rightarrow \infty}\left[\frac{r^{\Delta } \left((\Delta-3) \, \Phi (r)-r\,\Phi'(r)\right)}{2 \Delta -3}\right].
\end{align*}
Now, the constants $O_s$ and $O_c$ can be computed numerically. In order to solve the equation (\ref{sceqn}) we further expand $\Phi(r),f(r)$ near the horizon $r_0$ as
\begin{align*}
 &\Phi(r)\approx\Phi(r_0)+(r-r_0)\Phi'(r_0)+\frac{(r-r_0)^2}{2!}\Phi''(r_0)+\cdots,\\
 &f(r)\approx f(r_0)+(r-r_0)f'(r_0)+\frac{(r-r_0)^2}{2!}f''(r_0)+\cdots.
\end{align*}
Using the above series expansion into the equation (\ref{sceqn}) and demanding the regularity condition at the horizon, one obtains following constraints on the field at the horizon, 
\begin{align*}
\Phi'(r_0)=\frac{(\Phi(r_0)^2+m_{\phi}^2L^2)}{r_0^2f'(r_0)}\Phi(r_0).
\end{align*}
Thus, the choice of $\Phi(r_0)$ will automatically fix $\Phi'(r_0)$, and we can obtain a complete solutions for $\Phi(r)$. As discussed we have considered two possibilities. By shooting from the horizon with the scalar field value $\Phi(r_0)$, we studied sourceless condition $O_s=0$, which happens only below a critical temperature. For the other case, we studied with boundary source $O_s\neq 0$, at arbitrary temperature.
Before we move on, let us understand the behaviour of the scalar field and its dual nature with respect to the near horizon CFT. We examine the AdS$_2$ behaviour of the scalar field in the limit $r\rightarrow\, r_0$ and $T\rightarrow\,0$\ by writing the scalar field as
\begin{equation}
\Phi=\Phi(r_0)\,+\,\Phi_1\,(r-r_0)^\delta .
\end{equation}
It is the value of $\delta$ which controls the conformal dimension of the IR CFT dual operator. Therefore, by plugging into the equation of motion and solving the coefficients, we get 
\begin{equation}
\delta_{\pm}\,=\,-\frac{1}{2}\,\pm\,\sqrt{\frac{1}{4}+L_2^2\left(3\Phi(r_0)+m_{\Phi}^2\right)}
\end{equation} 
with $L_2$ being the AdS$_2$ radius. Condition for the dual of the scalar field to be an irrelevant deformation \cite{Iqbal:2010eh} in the IR CFT is $\delta=\delta_+>0$. This is also consistent with the scalar field which is going to be constant, $\Phi(r_0)$, in the near horizon limit. Therefore, under this neutral scalar field, properties of the back reacted near horizon geometry will remain same. A more elaborate discussion on this back-reaction issue can be found in \cite{Iqbal:2010eh}. We have also checked this by our full numerical solution. Therefore, we will mainly focus on the probe limit of the scalar field which makes our discussion simpler towards understanding the main goal of our paper. However, at this point we should mention that for a charged scalar field condensation such as for the holographic superconductor, the back-reaction near zero temperature has been proved to be very much important \cite{robert}. To this end let us emphasis again the reason for considering the scalar field. We will consider two possible scenarios while calculating the Fermionic spectral function. 

Case-I: For this case, we will consider the effect of scalar condensation on the Fermi surface. Therefore, we will consider those solutions of the scalar field, for which the boundary source $O_s$ will be zero. By appropriately tuning the background temperature and the horizon value of the scalar field $\Phi(r_0)$, one gets the desired solution with the condition $O_s=0$ and $O_c\neq 0$. In the left panel of Fig.\ref{fig1}, we showed the conventional condensation of the boundary dual scalar operator below the critical temperature $T_c$. As mentioned, we study how this temperature dependent scalar field profile controls the properties of low energy behaviour of the holographic Fermions at finite temperature. Our numerical computation gives $T_c \approx 0.001078$. What we will observe is that below this critical temperature, the holographic Fermi surface develops a pseudo-gap phase in its spectral function. It would be interesting to investigate the connection between our phase transition temperature $T_c$ with the well known cross over temperature $T^*$ in the high temperature superconductivity phase diagram. We have plotted the behaviour of the scalar field at the horizon with temperature in the right panel of Fig.\ref{fig1}.
Very close  to $T$=$T_c$ by fitting $O_c$ with $\delta_1(T_c-T)^{\delta}$, we found that the exponent $\delta=0.49\pm0.005$ and the proportionality constant $\delta_1$ is found to be $\delta_1\approx 1.41$. Similarly, the behaviour of $\Phi(r_0)$ $vs$ $\alpha_1(T_c-T)^{\alpha}$ gives the constant $\alpha_1=33.24$ and the exponent $\alpha$ to be $0.53\pm0.02$. 

Case-II: For this case we will identify $O_s$ as a tuning parameter and our goal would be to study the  effect of this source on the boundary Fermionic spectrum. Therefore, this source can be thought of as doping in the higher temperature superconductivity phase diagram. In Fig.\ref{fig2}, we show how the boundary source changes depending upon the horizon value the bulk scalar field $\Phi(r_0)$. As it is the horizon value of the scalar field which controls the Fermion-gauge coupling in the bulk spacetime, study of the effect of the source as a tuning parameter, on the boundary Fermionic spectrum is an important aspect.\par

In the next subsection we shall examine the effects of this scalar field in the Fermion spectral function and see how the Fermi surfaces  and Fermi arcs evolve across the phase transition. 

\section{Fermion Lagrangian and Dirac equation}
In holographic approach to Fermionic systems, there can be a number of ways in which Fermions are coupled to gravity and gauge fields. One such non-minimal coupling is dipole-coupling, which was introduced in \cite{Edalati:2010ww}. Coupling between Fermions with background condensation has been studied in \cite{Faulkner:2009am}. In this work, the authors discuss effects of a superconducting condensate on holographic Fermi surfaces. They found stable quasiparticles with a gap using coupling between the Fermion and condensate.  In \cite{Edalati:2010ww}, Fermions coupled to gauge fields via a dipole interaction in the bulk was studied. By varying the strength of the interaction, it was shown that a new band in the density of states can be generated where the spectral density is transferred between bands.  Beyond a critical interaction strength, a gap opens up dynamically. The effects of a superconducting condensate on the holographic Fermi surfaces was studied in \cite{Faulkner:2009am}. Choosing a suitable coupling between the Fermion and the condensate, the work had shown that there exists stable quasi-particles with a gap. Further, finding similarities with the behaviour of the cuprates high temperature superconductor, it was found that a stable quasiparticle peak can appear in the condensed phase in the systems under study whose normal state is a non-Fermi liquid with no stable quasiparticle. Fermions were also studied in an electrically-probed and asymptotically AdS-Schwarzschild spacetime \cite{Vanacore:2015poa}. The dual Fermion two-point correlator was computed and the bulk interactions were shown to create anisotropic gaps in the Fermi surfaces of the boundary spectrum. Consequently, the chiral symmetry breaking Pauli coupling provided a holographic model for Fermi arcs. \par 
Our goal of the present paper would be to control the  aforementioned coupling by a scalar field through condensation or by tuning the dual operator at the boundary.
We will consider two different models as discussed above. The motivation to consider two different Fermion-scalar models is to understand better the mechanism of observing the pseudo gap in the Fermion spectral function. Further, it is an important question and still a debated issue in condensed matter physics that whether the Fermi arcs arise due to the partial gapping of the Fermi surface or due to the certain destruction of quasi-particles.
In the holographic framework we will try to understand this question for both models. 
\subsection{Model-A}
Firstly, we generalise  \cite{Vanacore:2015poa} with similar action given by
	  \begin{equation}\label{model1}
	  \mathcal{S}^{(1)}_{Fermion}=\int d^4x\sqrt{-g}i\bar{\psi}\left(\slashed{D}-m-ip\Gamma\Phi\slashed{F}\right)\psi
	  \end{equation}
	  where the matrix factor, $\Gamma=\gamma^{\underline{r}}\gamma^{\underline{t}}(\hat{n}\cdot\vec{\Gamma})$.
	  Dirac equation of motion for $\psi$ is 
	  \begin{equation}\label{diraceqn}
	  \left(\slashed{D}-m-ip\Gamma\Phi(r)\slashed{F}\right)\psi=0 
	  \end{equation}
	  with  $\vec{\Gamma} \equiv (\Gamma^{\underline{x}}, \Gamma^{\underline{y}})$. By choosing $\hat{n} = \hat{x}$ 
	  and the following form of the Dirac matrices
	  \begin{align}\label{gammamatrices}
	  &\gamma^{\underline{r}}=\left( \begin{array}{ccc}-\sigma_3 & 0  \\0 & -\sigma_3 \end{array}\right),
	  &\gamma^{\underline{t}}=\left( \begin{array}{ccc}i\sigma_1 & 0  \\0 & i\sigma_1 \end{array}\right)\nonumber \\
	  &\gamma^{\underline{x}}=\left( \begin{array}{ccc}-\sigma_2 & 0  \\0 & \sigma_2 \end{array}\right),
	  &\gamma^{\underline{y}}=\left( \begin{array}{ccc}0 & \sigma_2  \\\sigma_2 & 0 \end{array}\right)
	  \end{align}
	  
	  For non trivial profile of the scalar field $p_{eff} = p \Phi$ plays the role of anisotropic coupling between Fermion and gauge field. By choosing the following ansatz for the Fermion field  $\psi(r,\vec{x_i})=(-gg^{r r})^{-\frac{1}{4}}e^{-i\omega t+i k.x}\tilde{\psi}(r,k)$, one can get rid of the spin connection, and finally the Dirac equation (\ref{diraceqn}) transforms into,
	  \begin{align}
	  &\Big[\frac{1}{\sqrt{g_{rr}}}\gamma^{\underline{r}}\partial_r+\frac{1}{\sqrt{-g_{tt}}}\gamma^{\underline{t}}\left(-i\omega-iqA_t\right)+\gamma^{\underline{x}}\frac{ik_x}{\sqrt{g_{xx}}}+\gamma^{\underline{y}}\frac{ik_y}{\sqrt{g_{yy}}}\nonumber\\
	  &-m-ip\Phi\frac{\gamma^{\underline{r}}\gamma^{\underline{t}}\gamma^{\underline{x}}}{\sqrt{-g_{tt}g_{rr}}}\gamma^{\underline{r}}\gamma^{\underline{t}}\partial_rA_t\Big]\tilde{\psi}=0
	  \end{align}
	  Considering the following component form of the Fermion field,  $\tilde{\psi}=\left(\tilde{\psi}_1,\tilde{\psi}_2\right)^T$, and ansatz for the background gauge field $A_{\mu}=A_t$ with all other components to zero, above equation can be further simplified to
	  \begin{align}\label{diraceqn2}
	  \frac{1}{ \sqrt{g_{rr}}}\partial_r\left(\begin{array}{c}\tilde{\psi}_1  \\ \tilde{\psi}_2  \end{array}\right)&=\frac{1}{\sqrt{-g_{tt}}}(\omega+qA_t)i\sigma_2\otimes\left(\begin{array}{c}\tilde{\psi}_1  \\ \tilde{\psi}_2  \end{array}\right)-m\sigma_3\otimes\left(\begin{array}{c}\tilde{\psi}_1  \\ \tilde{\psi}_2  \end{array}\right)\mp \frac{k_x}{\sqrt{g_{xx}}}\sigma_1\otimes\left(\begin{array}{c}\tilde{\psi}_1  \\ \tilde{\psi}_2  \end{array}\right)\nonumber\\
	  &\pm \frac{p\Phi}{\sqrt{-g_{tt}g_{rr}}}A_t'\sigma_1\otimes\left(\begin{array}{c}\tilde{\psi}_1  \\ \tilde{\psi}_2  \end{array}\right)+\frac{k_y}{\sqrt{g_{yy}}}\sigma_1\otimes\left(\begin{array}{c}\tilde{\psi}_2  \\ \tilde{\psi}_1  \end{array}\right)
	  \end{align}
	  The above equation (\ref{diraceqn2}), after inserting the background geometry (\ref{metric}) becomes
	  \begin{align}\label{diraceqns}
	  &r^2 \sqrt{f(r)} \partial_r \tilde{\psi}_{I} = \frac{i \sigma_2}{\sqrt{f(r)}} \left( \omega + q \mu \left(1- \frac{r_0}{r} \right) \right) \tilde{\psi}_{I}\nonumber\\& - \sigma_3  m r  \tilde{\psi}_{I} - (-1)^I \sigma_1 \left( p\Phi(r) \mu \frac{r_0}{r} - k_x \right) \tilde{\psi}_{I} + \sigma_1 k_y \tilde{\psi}_{J}.
	  \end{align}
	  The asymptotic solutions of (\ref{diraceqns}) are given by
	  \begin{align*}
	  \tilde{\psi}_{J}=A_J(k) r^{m}+B_J(k)r^{-m}
	  \end{align*}
	  In fact, we can numerically read off the coefficients $A$ and $B$ to obtain the green's functions, but alternative approach exists by solving the first order flow equation.\par
	  In equation (\ref{diraceqns}) we see that with $k_y\ne 0$ the block diagonal form is lost, we now have mixing of various spinors components. Following the prescription  used in \cite{Faulkner:2009am,Guarrera:2011my, Vanacore:2015poa} to extract the Green's
	  function by using two sets of linearly independent boundary conditions are given by
	  \begin{align}\label{defn}
	  \left(\begin{array}{cc}\beta^I_1& \beta^{II}_1\\ \beta^I_2 & \beta^{II}_2 \end{array} \right)=  \left(\begin{array}{cc} s_{11}& s_{12}\\ s_{21} & s_{22} \end{array} \right) \left(\begin{array}{cc}\alpha^I_1& \alpha^{II}_1\\ \alpha^I_2 & \alpha^{II}_2 \end{array} \right),
	  \end{align}
	  where we further expressed two component spinor as,  $\tilde{\psi}_i=\left(\beta_i,\alpha_i\right)^T$. The retarded Green's function is defined as   
	  \begin{align}\label{GreenFunction}
	  G_R(\omega,\vec k)=-i \left(\begin{array}{cc} s_{11}& s_{12}\\ s_{21} & s_{22} \end{array} \right)\cdot \gamma^t,
	  \end{align}
	  with gamma matrices  $\gamma^t=i\sigma_1$. The spectral function is defined as
	  \begin{align}
	  A(\omega,\vec k)=Im\left[Tr G_R(\omega,\vec k)\right].
	  \end{align}
	  Along with the definition given in (\ref{defn}) and from equation (\ref{diraceqn2}), we can derive the  flow equation  given by
	  \begin{align}\label{flow1}
	  \frac{1}{\sqrt{g_{rr}}}\partial_rG_R+2mG_R=M_+-G_RM_-G_R
	  \end{align}
	  where,
	  \begin{align*}
	  M_{\pm}=\left(\begin{array}{cc}
	  \pm W_{\pm}-\frac{k_x}{\sqrt{g_{xx}}} & \frac{k_y}{\sqrt{g_{yy}}}  \\
	  \frac{k_y}{\sqrt{g_{yy}}} &\pm W_{\mp}+\frac{k_x}{\sqrt{g_{xx}}}
	  \end{array}
	  \right)
	  \end{align*}
	  with $W_{\pm}$  given by 
	  \begin{align}
	  W_{\pm}=\frac{1}{\sqrt{-g_{tt}}}\left(\omega+qA_t \right)\pm \frac{p\Phi }{\sqrt{\-g_{tt}g_{rr}}}A_t' .
	  \end{align}
	  Numerically we will integrate equation (\ref{flow1}) from horizon ($r=r_0$) to infinity in order to compute the spectral function.The boundary condition for $\omega\ne 0$ is still in diagonal form given by
	  \begin{align}
	  G_R(r_0)=\left(\begin{array}{cc}
	  i & 0 \\
	  0 &i
	  \end{array}
	  \right) .
	  \end{align}
	  
	 \subsection{Model-B}
Even though most of our discussions will be focused on the previous model, for completeness and comparison specifically in the context of energy gap in the spectral function, we consider the following dipole model generalising the work of \cite{Edalati:2010ww} by coupling a scalar field as part of the controlling parameter
\begin{equation}\label{model2}
 \mathcal{S}^{(2)}_{Fermion}=\int d^4x\sqrt{-g}i\bar{\psi}\left(\slashed{D}-m-ip\Phi\slashed{F}\right)\psi,
\end{equation}
where,\begin{align}\slashed{D}=&e^{\mu}_c\gamma^c\left(\partial_{\mu}+
\omega_{\mu}^{ab}-iq A_{\mu}\right)\nonumber\,\,\,,\,\,
\slashed{F}=\frac{1}{2}\gamma^{ab}e^{\mu}_ae^{\nu}_bF_{\mu\nu} .
 \end{align}
 The parameter $p$ is a Pauli coupling, $e^{\mu}_a ,\omega_{\mu}^{ab}$ are vielbeins and spin connection. Here, $\{a,b\}$ are tangent space indices and $\{\mu,\nu\}$ are for the bulk. As shown in  appendix \ref{appa}, the Green's function is given by
\begin{align}
G_R(\omega,k)=\lim_{r\rightarrow \infty}\frac{1}{r^{2m}}\left( \begin{array}{ccc}\zeta_+ & 0  \\0 & \zeta_- \end{array}\right)
\end{align}
 The spectral function defined as 
\begin{equation}
A(\omega,k_x)\,=\,Tr \left[Im \left(G_R(\omega,k_x)\right)\right].
\end{equation}
For AdS$_2$ Green's function we refer to appendix \ref{appb}, which includes the finite temperature dependent scalar field and how it changes the IR CFT operators.


\section{Numerical results and discussions}

\subsection{Across the phase transition: without source}

As discussed above, we have two different kind of solution for the scalar field. In this subsection we will discuss case-I, when the scalar field condenses below a critical temperature $T_c$. Therefore, above $T_c$ we will have free Fermion with well defined Fermi surface. However below $T_c$ we will have non-trivial properties of the Fermion spectral function. In the left panel of Fig.\ref{fig4}, we see the evolution Fermi surface from higher temperature to lower temperature across the critical temperature $T_c $. For $T>T_c$, scalar field does not condensate leading to the closed Fermi surface. We lowered the temperature, consequently, Fermi surface starts to develop anisotropic gap which we call pseudo-gap. Important to mention that even though our results may look  similar to the one shown in \cite{Vanacore:2015poa}, it is the evolution of Fermi arc with respect to the temperature below a critical value, which can be identified with the pseudo-gap region at constant doping for the cuprate superconductor. Important difference is the arc topology of the holographic Fermi arc compared to the d-wave symmetric Fermi arc of the real high temperature superconductor. Therefore, we need to further work on the issue of understanding the d-wave symmetric Fermi arc for our holographic system.  Nonetheless, an important point to emphasise that understanding the pseudo-gap phenomena is still an active area of condensed matter research. It is interesting to re-emphasise that the holographic dual of our bulk scalar field can be interpreted as an incoherent phase fluctuation, which was proposed as a potential mechanism for the pseudo-gap phenomena \cite{PhysRevB.54.R3756}. We also plotted in fig.\ref{fig4} the evolution of Fermi surface for higher $p$ value which essentially changes the absolute strength of the gauge-Fermion coupling. We will further explore on this in the future publication. From pole/zero duality perspective \cite{PhysRevD.90.126013,Ling2014}, the appearance of the gap can be realised from the pole and zero in $G_{11}\,\text{and}\,G_{22}$ which are the diagonal components in the Green's function. The effective coupling $p\Phi$ and the $\Gamma$ matrix in the $\hat{n}$-momentum axis inverts the sign of Fermi momentum $k_f$ for negative and positive $p$ values. As a results of this the gap appear in either $-\bold{k}_f$ or at $\bold{k}_f$. As the Fermi surface is anisotropic, we now investigate how the magnitude of the Fermi momentum $k_f$ changes as we go along the surface. 

\begin{figure}[h]
	\centering
	\includegraphics[scale=0.7]{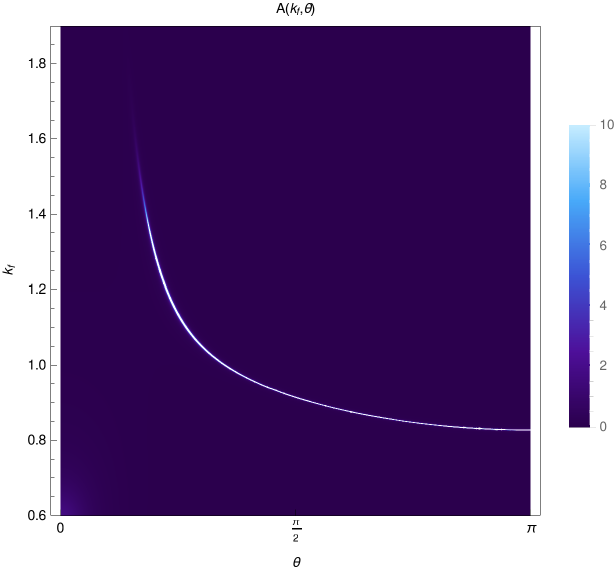}
	\caption{Plot of $A(k_f,\theta)$ in $\theta-k_f$. Here $p=1.5$\hspace{0.1cm},$q=1$,\hspace{0.1cm}$m=0$,\hspace{0.1cm}$m_{\Phi}^2=-21/10$ and at very small $T=10^{-3}T_c$. In this plot we see the variation of $k_f$ along $\theta$ direction.}
	\label{fig3}
\end{figure} 
\begin{table}[h] 
	\begin{center}
		\caption{ Variation of $k_f$ for different $\theta$ with p=1.5 and q=1.}
		\label{tab1}
		\vspace{0.2cm}
		\begin{tabular}{ p{1.5cm}p{2cm}p{2cm} }
			\hline
			&$k_f$  & $\theta$\\
			\hline\vspace{0.1cm}\\
			&$\approx$1.5240 & $\frac{\pi}{6}$\vspace{0.1cm} \\
			& $\approx$1.0150 & $\frac{\pi}{3}$\vspace{0.1cm} \\ 
			 &$\approx$0.9146 & $\frac{\pi}{2}$\vspace{0.1cm} \\ 
			&$\approx$0.83633 & $\frac{5\pi}{6}$\vspace{0.1cm} \\ 
			&$\approx$0.86388 & $\frac{2\pi}{3}$ \vspace{0.1cm}\\ 
			&$\approx$0.84776 & $\frac{3\pi}{4}$ \vspace{0.2cm}\\ 
			\hline
		\end{tabular}
	\end{center}
\end{table}

 In order to find $k_f$ in $k_x$-$k_y$ plane we will define $k_x,k_y$ in terms of angle $\theta$ in [$k_x$-$k_y]$-plane as
 \begin{align}
  &k_x=k_f\hspace{0.1cm} \sin \theta ;\hspace{0.5cm}k_y=k_f\hspace{0.1cm} \cos \theta ,
  \label{kfdef}
 \end{align}
where $k_f$ is the distance of the Fermi surface from the centre. In Table-\ref{tab1} we tabulate different values of $k_f$ in different directions($\theta $ values) in the $k_x$-$k_y$ plane. Numerically we found that $k_f$ is same for all angular coordinate for $T>T_c$. This is expected from the fact that in this limit the anisotropic gauge field and Fermion coupling vanishes because of zero scalar field value. With the definition in equation (\ref{kfdef}), we plotted the spectral function in  Fig.\ref{fig3}. 
 
 \subsection{At arbitrary temperature: with source}
For case-II, as mentioned earlier we will consider the bulk scalar field with non-normalizable solution. Therefore, we have only one tuning  parameter corresponding to the source of the dual operator which we tune to evolve the Fermi surface at a particular temperature. 
 \begin{figure*}[h]
 	\centering
 	\begin{minipage}{.5\textwidth}
 		\centering
 		\includegraphics[scale=0.40]{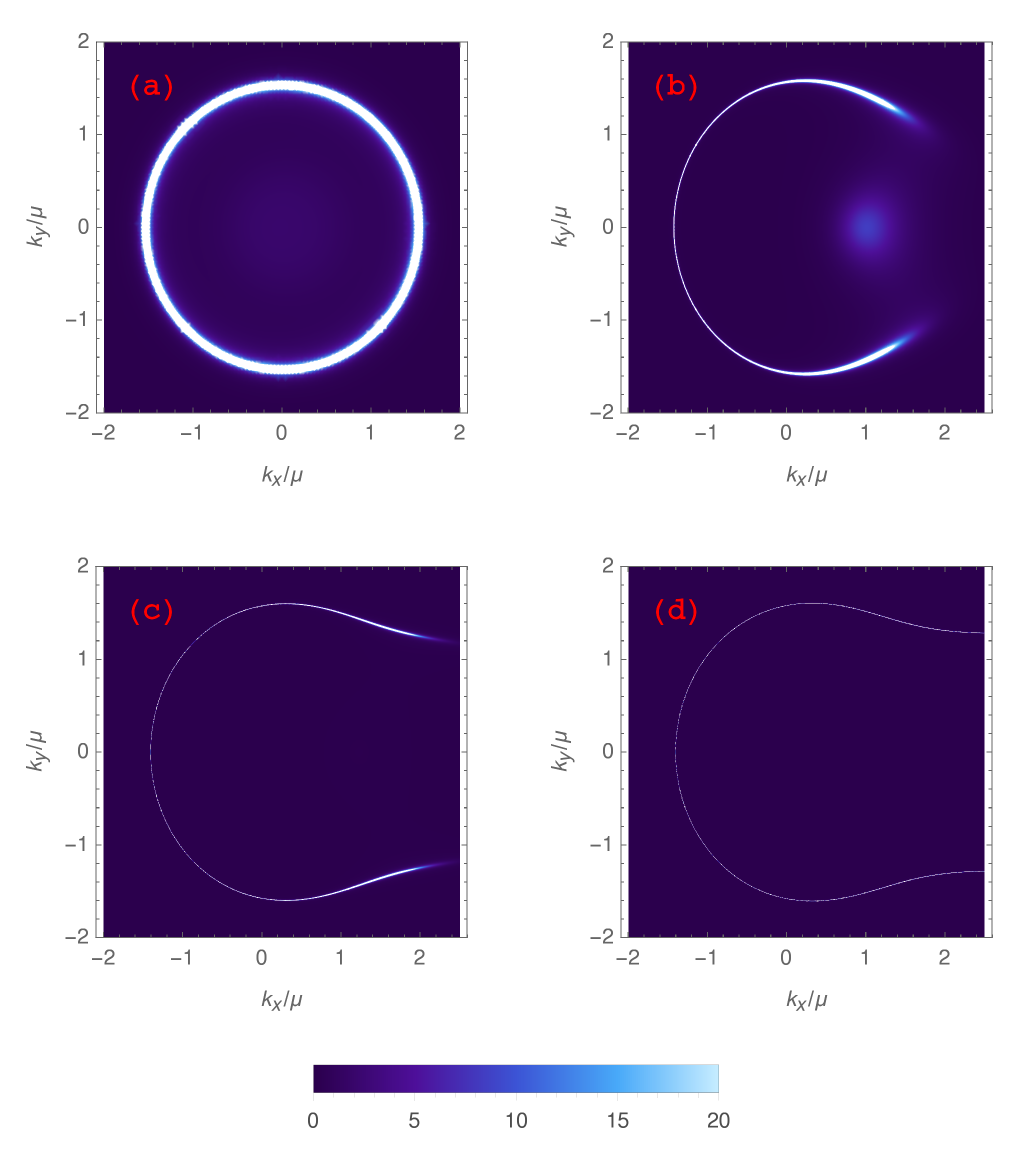}
 	\end{minipage}%
 	\begin{minipage}{.5\textwidth}
 		\centering
 		\includegraphics[scale=0.31]{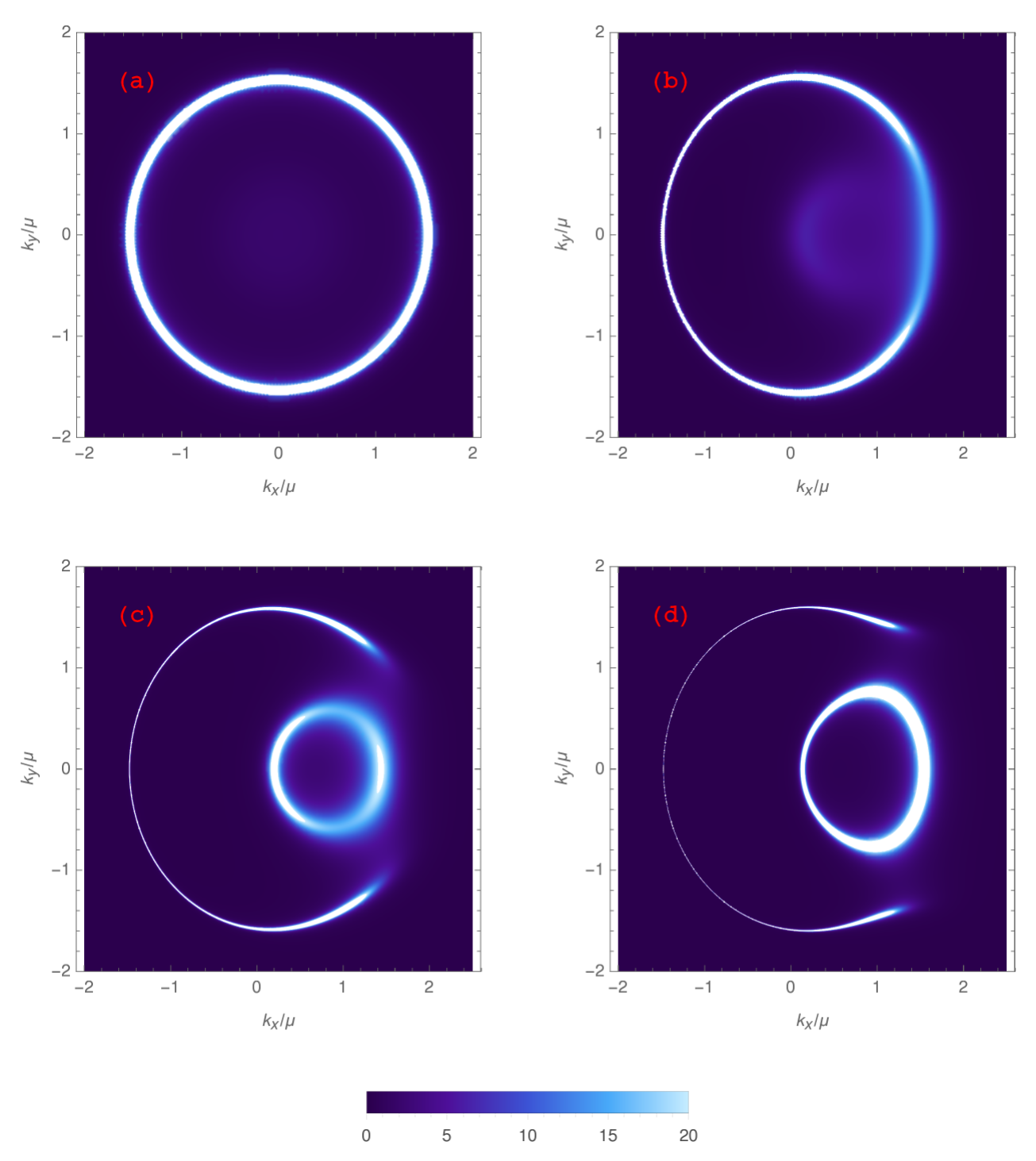}
 	\end{minipage}
 	\caption{Density plot of spectral function $A(k_x,k_y)$ with small $\omega$ (=$0.0001$), $q$=1 and fixed $p$=2 (Left) $p$=0.92 (Right). The temperature of above plots from (a)-(d) are $ 0.99T_c,0.55T_c,0.18T_c \text{ and } 10^{-3}T_c$ respectively.}
 	\label{fig4}
 \end{figure*}
 \begin{figure*}[h]
 	\centering
 	\begin{minipage}{.5\textwidth}
 		\centering
 		\includegraphics[scale=0.8]{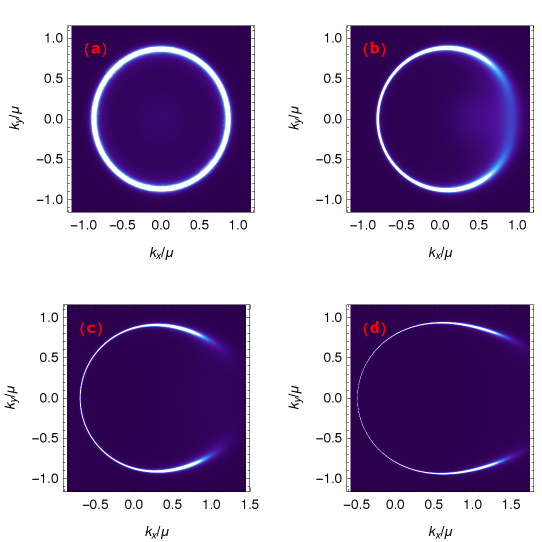}
 	\end{minipage}%
 	\begin{minipage}{.5\textwidth}
 		\centering
 		\includegraphics[scale=0.8]{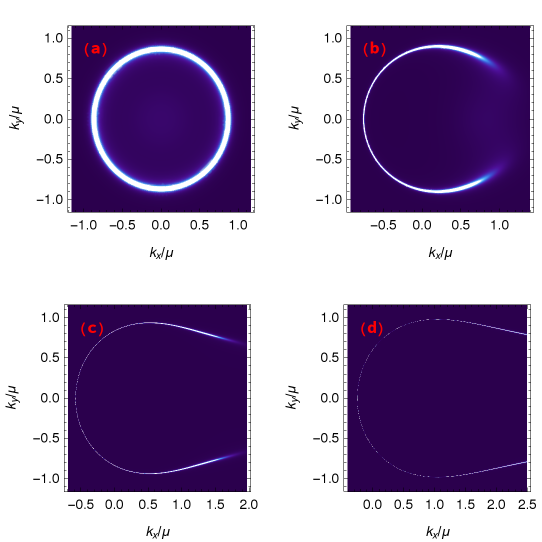}
 	\end{minipage}
 	\includegraphics[scale=0.6]{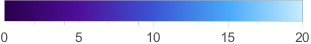}
 	\caption{{Plot of spectral function $A(k_x,k_y)$ for fixed temperature $T\approx 0.00238$, $q$=1 for $p=1$(left) and $p=2$(right). From [a-d] corresponds to $\Phi(r_0)= 0,0.5,1 \text{ and }1.3$   respectively. }                                              }
 	\label{fig5}
 \end{figure*} 
We have chosen two different value of $p$ and the plots for the spectral function are shown in the Fig.\ref{fig5}. For this particular case we do not have any critical temperature. For a fixed temperature of the system the Fermi surface evolves as we tune the boundary source, which essentially control the scalar field profile in bulk. For zero source we will obviously have closed spherical Fermi surface. In contrast to the previous case-I, the boundary tuning parameter of present case can be identified with the doping in the high temperature superconductor. Qualitatively we do not find much difference between the Fermi surface properties with increasing $p$ as is clear from the Fig.\ref{fig4}.

\section{Energy gap in the spectral function}
\begin{figure*}[h]
	\centering
	\includegraphics[scale=0.5]{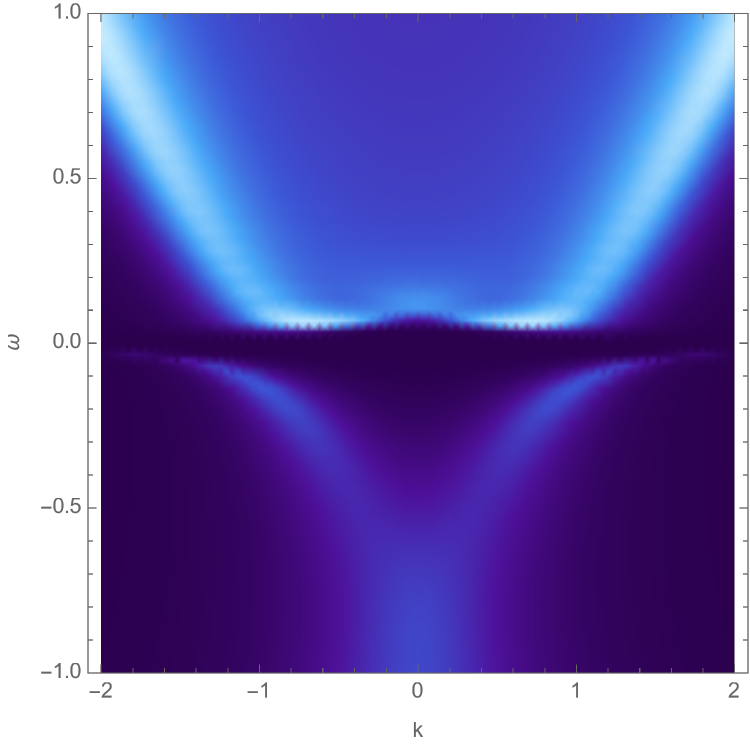}
	\includegraphics[scale=0.5]{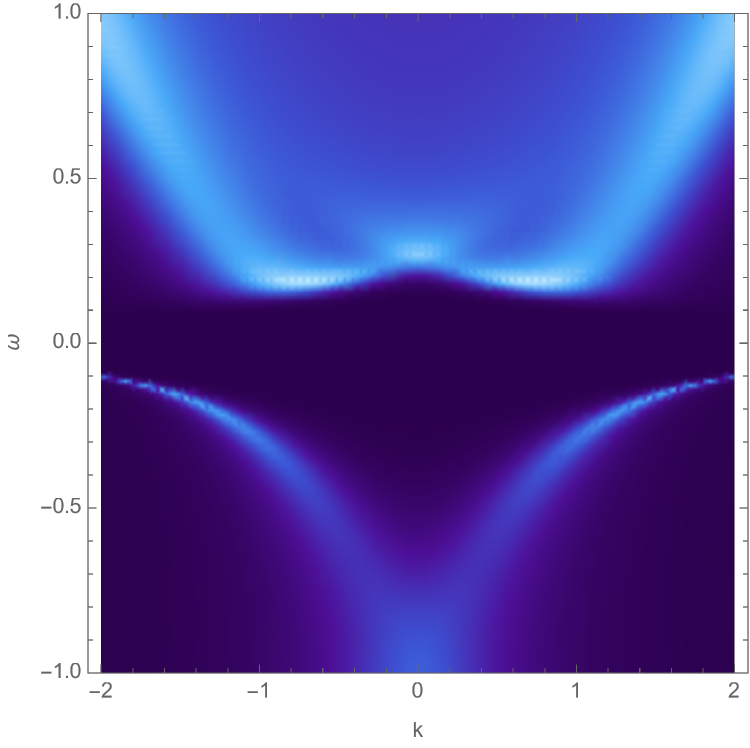}
	\hspace{0.4cm}
	\caption{Spectral function $A(\omega,k)$ vs ($\omega,k$) for model-B. Below $T_c$ ($ \approx 0.001078$), the opening of gap near $\omega=0$ is seen along $k_x$ direction for fixed $p=5\,(left)$ and $p=10\,(right)$, whereas above $T_c$ the gap disappeared\,.\,}
	\label{fig6}
\end{figure*}

\begin{figure*}[t]
	\centering
	\includegraphics[scale=0.5]{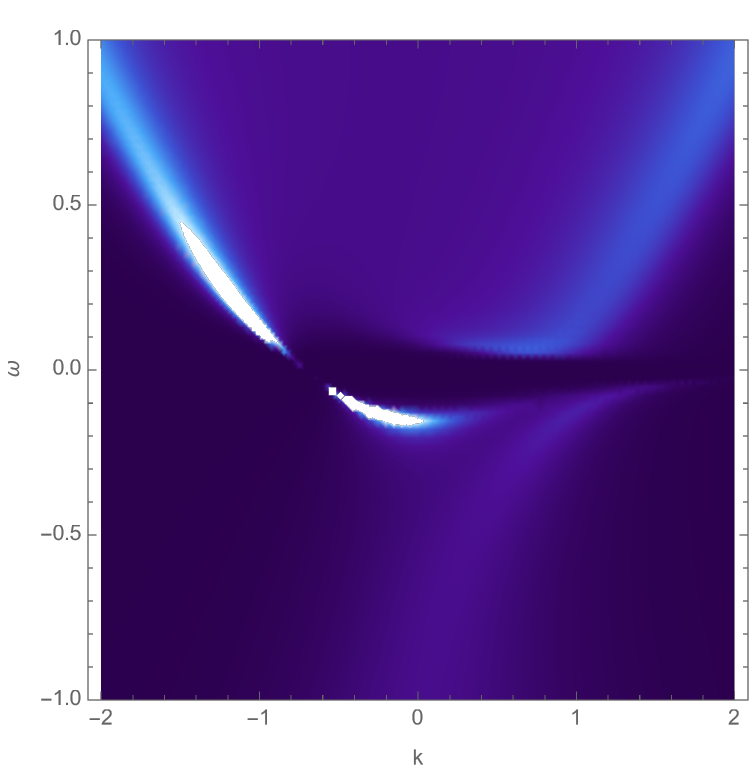}
		\includegraphics[scale=0.5]{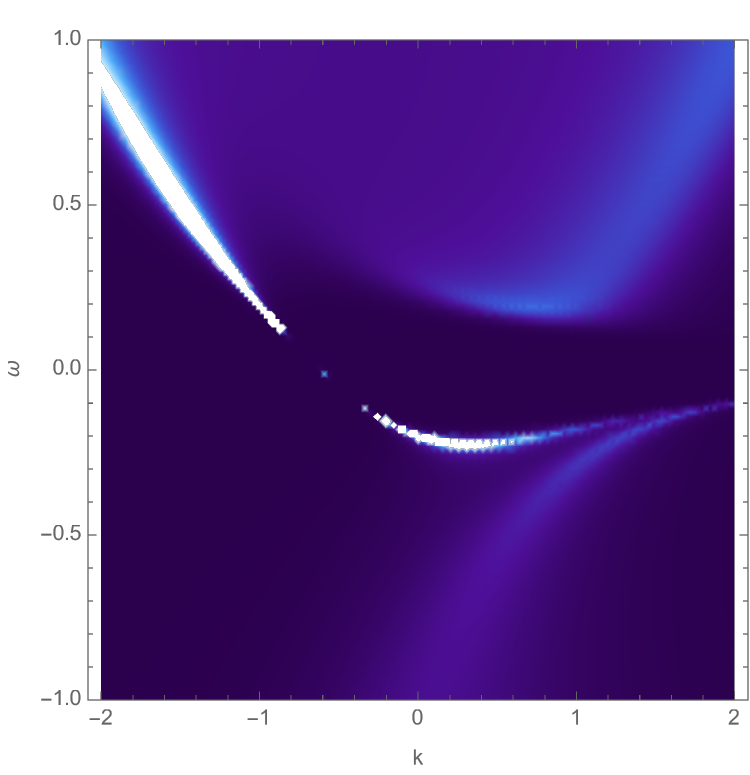}
	\hspace{0.2cm}
	\caption{Spectral function $A(\omega,k)$ vs ($\omega,k$) for action \ref{model1} for model-A. Below $T_c$, the opening of gap near $\omega=0$ is seen along $k_x$ direction for fixed $p=5\,(left)$ and $p=10\,(right)$.}
	\label{fig7}
\end{figure*}

In this section we consider an important question related to the existence of actual gap in the Fermion spectral function. As already pointed out before, it is still a debated issue from the experimental point of view whether the Fermi arcs arise due to the partial gapping of the Fermi surface or due to the certain destruction of quasi-particles. Therefore, it is important to examine the situation for our holographic set up. We consider two different models-(A,B). 
Figure \ref{fig6} illustrates the presence of the gap for model-B, which contains dipole type Fermion-gauge interaction with scalar field dependent effective coupling parameter $p_{eff}= p\Phi$. For the usual no-scalar field dipole model described in  \citep{Edalati:2010ww} the energy gap has already been found out to be much wider than the present model. However, clearly the size of the gap increases with increasing $p$ value. At this point, it is also important to note the symmetric nature of the energy gap for the dipole type coupling. On the other hand for the model-A, because of the parity breaking Fermion-gauge coupling, we have already seen the anisotropic Fermi arc Fig.\ref{fig4}, and that is indeed found to be translated into the anisotropic energy gap in the spectral function shown in Fig.\ref{fig7}. For this plot we set $k_y=0$ and $k_x=k$.
Therefore, in the positive $k$ direction we can clearly see the gap whose width is monotonically increasing with $p$ value. Thus, in holographic pseudo-gap phase, the Fermi arcs seem to be intimately connected with the partial gapping of the Fermi surface. It would, therefore, be interesting to construct a holographic model where partial gapping is not occurring.  


In our subsequent section we will try to understand the low energy properties of the Fermi surface for both the cases by considering the well known analytic technique in terms of $AdS_2$ Fermionic spectral function.    

\section{Analytical study of Green's function at finite temperature}
In this section we will try to understand how the scalar field enters into the Fermi arc dynamics using analytic treatment. We essentially follow the work of \cite{Faulkner:2009wj,Faulkner:2011tm} and pin point the contribution of the scalar field contribution. In the AdS$_2$ limit  at temperature $T\,\rightarrow\,0$, Eq.\ref{diraceqns} takes the form (see appendix \ref{appc})
\begin{align}
-\zeta\partial_{\zeta}\tilde{\psi}_{I} =i\sigma_2\left(\omega\zeta+q e_d\right)\tilde{\psi}_{I} -\sigma_3mL_2\tilde{\psi}_{I}-(-1)^I\sigma_1L_2m_k\tilde{\psi}_{I} +\sigma_1L_2k_y\tilde{\psi}_{J} 
\end{align}
For finite temperature, the boundary Green's function is expressed as \cite{Faulkner:2013bna}  

\begin{equation}
 G_R=\frac{\tilde{B}_+ +\mathcal{G}_RT^{2\nu_k}\tilde{B}_-}{\tilde{A}_+ +\mathcal{G}_RT^{2\nu_k}\tilde{A}_-},
\end{equation}
where, $\tilde{B}_{\pm}$ and $\tilde{A}_{\pm}$ are all matrices. For small $\omega$ one can perturbatively expand $\tilde{B}_{\pm}$ , $\tilde{A}_{\pm}$ in terms of $\omega$.
Where, the finite temperature $AdS_2$ Green's function is $\mathcal{G}_R$ 
\begin{align}\label{gtgreen}
\mathcal{G}_R(\omega,T)=c_k\,\,(4\pi T)^{2\nu_k}
\end{align}
where, $c_k$ is a normalisation constant and $ \nu_k= \sqrt{m^2L_2^2+m_k^2L_2^2+k_y^2L_2^2-q^2e_d^2}$ .
Here, $\nu_k$ plays the role of conformal dimension of dual infrared conformal field theory operator. $L_2 = {1}/{\sqrt{6}}$ is the $AdS_2$ radius in unit of $AdS_4$ radius $L$, and $m_k=\left(\Phi(r_0)\mu-k_x\right)$. We have already observed through our numerical calculation the anisotropic and temperature dependent behaviour of the Fermi surface. It is important to see that those behaviour is manifested into the expression for the conformal dimension $\nu_k$ through the scalar field condensation $\Phi(r_0)$, and component of Fermi momentum $(k_x,k_y)$.  
Obviously, at low energy the boundary Green's function will be mainly controlled by $\mathcal{G}_R$. Fermi surface is associated with the zeros of $\tilde{A}_{\pm}$. Therefore at small energy and momentum near the Fermi surface ($|\vec{k}|=|k_f|$) the Green's function can be expressed in the following form \cite{Faulkner:2013bna}
%
\begin{align}\label{aprxgreen}
 G_R=\frac{\tilde{H}_1}{k_{\perp}-\omega V_f^{-1}+D_3 T- \tilde{H}_2T^{2\nu_{k_f}}\mathcal{F}_{k_f}(\nu_{k_f},\frac{\omega}{T})},
\end{align}
where $\mathcal{F}$ is given by 
\begin{align*}
 \mathcal{F}_{k_f}=\frac{\Gamma(\frac{1}{2}+\nu_{k_f}-\frac{i\omega}{2\pi T}+iq e_d)}{\Gamma(\frac{1}{2}-\nu_{k_f}-\frac{i\omega}{2\pi T}+iqe_d)} .
\end{align*}
In (\ref{aprxgreen}), $\tilde{H}_1,V_f,\tilde{H}_2$ has the form as in \cite{Faulkner:2013bna} and $D_3$ is a constant which can be obtained by full numerical calculation.

{\em Zero temperature limit:} When $\frac{\omega}{T}\rightarrow\infty$(zero temperature), $\mathcal{F}$ behave as
\begin{align*}
\mathcal{F}\approx e^{-i \pi  v_{k_f}} \left(\frac{\omega}{T}\right)^{2 \nu_{k_f}},
\end{align*}
which gives us back the zero temperature $AdS_2$ Green's function.
The Fermi surface is defined by
the pole of $G_R$ expressed as
\begin{align*}
\det\left[k_{\perp}-\omega V_f^{-1}-\tilde{H}_2 e^{-i\pi \nu_{kf}}\omega^{2\nu_{k_f}}\right]=0
\end{align*}
The properties are usually measured by the dispersion relation near the Fermi surface which is the pole of the Greens's function,
\begin{align*}
\omega_p(k)&\equiv \omega_*(k)-i\Omega(k)
\end{align*} 
\begin{figure}[t]
	\centering
	\includegraphics[scale=0.8]{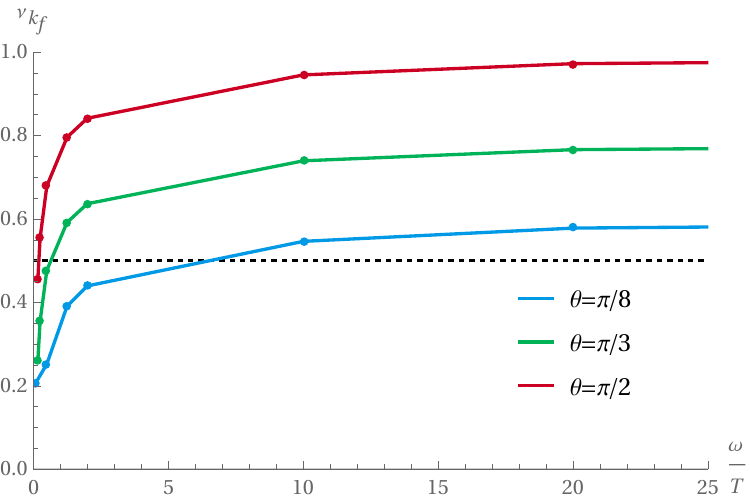}
	\caption{Plot of $\nu_{k_f}$ as a function of $\omega/T$ keeping $\omega$ fixed. Behaviour is in the condensed phase of the scalar field. We considered  $p=1,m=0,q=1$, dashed line corresponds to $\nu_{k_{f}}$=1/2. It illustrates how $\nu_{k_{f}}$ changes for $\frac{\omega}{T} \rightarrow\,\infty$ as seen in AdS$_4$ Green's function. As we decrease $T$ for fixed small $\omega$, $\nu_{k_{f}}$ decreases below $\nu_{k_{f}}=1/2$.}
	\label{fig8}
\end{figure}

\begin{figure}[t]
	\centering
	\includegraphics[scale=0.8]{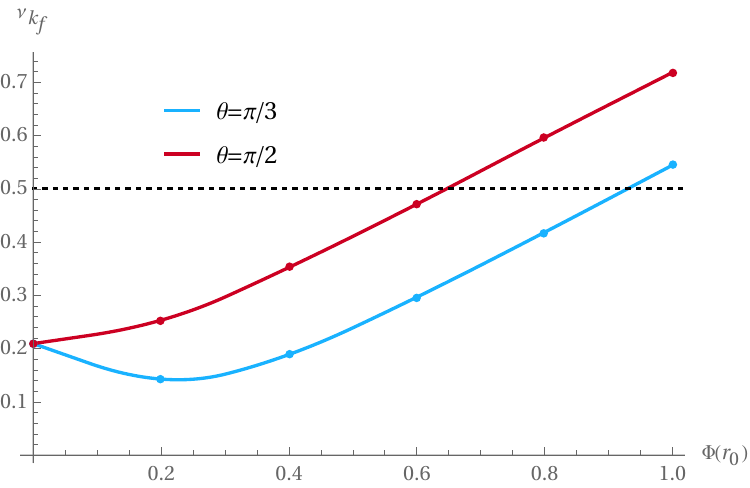}
	\caption{Plot of $\nu_{k_f}$ as a function of horizon value of the scalar field $\Phi(r_0)$ associated with non-zero source at the boundary. Temperature is fixed at $T=0.00238$ for $q=1$,$ m=0,q=1$ and dashed line is $\nu_{k_{f}}$=1/2. Here, we take $m_{\Phi}^2=-1.4$, which is above the BF-bound to avoid the condensation of the scalar field.}
	\label{fig9}
\end{figure}
The dispersion relations are parametrized as $\omega_*(k)\propto k_{\perp}^z$ and widths $\Omega_*(k)\propto k_{\perp}^{\alpha}$ \cite{Faulkner:2009wj}. The exponents are 
\begin{equation*}
z=\begin{cases}
\frac{1}{2\nu_{k_f}} & \text{for }\nu_{k_f}<\frac{1}{2}\\    
1 & \text{for }\nu_{k_f}>\frac{1}{2}
\end{cases}~;~      
\alpha=\begin{cases}
1  & \text{for }\nu_{k_f}<\frac{1}{2}\\    
2\nu_{k_f} & \text{for }\nu_{k_f}>\frac{1}{2}
\end{cases}
\end{equation*}
Our strategy will be the following: For the mechanism when the neutral scalar field condensates below the critical temperature which is very small, the black hole near horizon geometry can be approximated as $AdS_2$. Near zero temperature similar approximation can also be made for the other mechanism when the spectral function is controlled by the external source of dual scalar operator at the boundary. With this assumption in mind we consider the Green's function with the horizon value of the scalar field to be temperature dependent and show the evolution of Fermi arc with temperature. Considering mass of the Fermion to be zero, 
\begin{eqnarray}
\nu_{k_{f}} &=& \sqrt{{m_{k_f}}^2L_2^2+k_{y_f}^2L_2^2-q^2e_d^2} \\ \nonumber
&=& \sqrt{\frac{1}{6}\left[\Phi(r_0)\left(\Phi(r_0)\mu^2-2 \mu k_{f} \cos\theta\right)\right]+k_f^2-q^2e_d^2} 
\end{eqnarray}
It is clear and well known that $\nu_k =1/2$ encodes the properties of marginal Fermi liquid which has been phenomenologically introduced to describe the strange metal phase of cuprate. From the above expression of $\nu_{k_f}$, it is clear that it is the horizon value of the scalar field $\Phi(r_0)$ which controls the properties of the Fermion spectral function. When the scalar field condensates below a critical temperature, $\nu_k$ naturally depends on the temperature through the scalar field value $\Phi(r_0)$ at the horizon. Therefore, below $T_c$ as one decreases the temperature, $\Phi(r_0)$ increases and consequently the properties of the Fermi surface changes from Fermi liquid, $\nu>\frac{1}{2}$, with long-lived quasi-particles to non-Fermi liquid $\nu_k <\frac{1}{2}$
with well defined quasi-particles shown the figure \ref{fig8}. The anisotropic nature of the Fermi surface can also be clearly observed. In the Fig.\ref{fig9}, we have plotted ${\nu_{k_f}}$ in terms of explicit scalar field horizon value $\Phi(r_0)$ for a fixed temperature. For this case the horizon value depends on the external source at the holographic boundary. One can, therefore, clearly see the quantitative difference between the two mechanisms of controlling the Fermionic spectral function.

%


\section{Summary and Conclusions}

Phenomena of pseudo-gap is an interesting area of condensed matter field in high temperature superconductivity. In the context of holographic method, one interesting mechanism of obtaining pseudo-gap was first proposed by Vanacore et. al. \cite{Vanacore:2015poa}. A non-minimal coupling between the Fermion and gauge field has been introduced in the AdS bulk. By appropriately tuning this free coupling parameter $p$, anisotropic gap in the Fermion spectral function is generated. In this paper we introduce a real scalar field whose non-zero profile modifies the aforementioned Fermion-gauge interaction in terms of $p_{eff} = p\Phi$, and thereby controls the boundary Fermi surface. We have considered two possible scenarios of generating this coupling. In the first scenario, the scalar field in the bulk acts as a neutral order parameter field at the boundary which gives rise to a phase transition from normal phase to pseudo-gap phase at a critical temperature $T_c$. As emphasised before, this $T_c$  can be identified with the well known crossover temperature $T^*$ in the high-$T_c$ superconducting phase diagram, below which pseudo-gap appears.
For holographic Fermions, parameter $\nu_k$ plays the role of operator dimension in the $AdS_2$, which controls the behaviour of the low energy Fermions at the boundary. In our analysis, this $\nu_k$ depends on the scalar field value at the horizon.
This in turn makes $\nu_k$ temperature dependent. We therefore, obtain the characteristic changes of the Fermi surface while changing the temperature below $T_c$ shown in Fig.\ref{fig4}. Because of strong anisotropic nature of the Fermi surface, we also discussed how the Fermionic properties changes along the Fermi surface from normal ($\nu_k < 1/2$) to marginal $(\nu_k = 1/2)$ and marginal to non-Fermi liquid $(\nu_k > 1/2)$ for a given temperature shown in Fig.\ref{fig8}.    
In the second scenario, we tune the non-minimal coupling $p_{eff}$ by the dual boundary scalar operator as a source, which essentially corresponds to the non-normalisable solution of the bulk scalar field. Hence for this system we do not have any critical temperature. Therefore, we studied this case for a fixed temperature and tune the boundary source which can be identified with the doping in high temperature superconductor. However, detail study needs to be done to understand this identification.

Finally we examine an important question related to the existence of actual gap in the Fermion spectral function. We consider two different holographic models-(A,B), associated with two different Fermion-gauge coupling prescription. 
The figure \ref{fig6} and \ref{fig7} illustrate the presence of the gap for both the models. For dipole type coupling the energy gap is symmetric in nature Fig.\ref{fig6}, whereas for the other model it is anisotropic Fig.\ref{fig6} in accordance with the anisotropic Fermi arc Fig.\ref{fig4}. 
Therefore, in holographic pseudo-gap phase, the Fermi arcs seem to be intimately connected with the partial gapping of the Fermi surface. It would, therefore, be interesting to construct a holographic model where partial gapping is not occurring.  

\section*{Acknowledgement} The authors would like to thank unanimous referees for their very useful suggestions in order to improve the quality of the paper.  

\appendix

\section{AdS$_2$ for Fermi arcs model-A}\label{appc}
Since bulk scalar field condensates below the critical temperature which is very small, the black hole near horizon geometry can be approximated as $AdS_2$. With this assumption we derive the boundary Green's function considering the horizon value of the scalar field to be temperature dependent and show the evolution of Fermi arc with temperature. Therefore, we begin by expanding the equation (\ref{diraceqns}) in the near horizon limit and focus only the effects of scalar field. Near the horizon($r=1$), $f(r)\approx 6(r-1)^2$, $A_t\approx \mu(r-1)$, also we will set $L=1$. Thus we arrive with the 
 equation given by
 \begin{align}\label{ads2eqn}
 \chi\partial_{\chi}\psi_I&=L_2\left[\sigma_3m+(-1)^I\sigma_1 m_k\right]\psi_I-i\sigma_2\left(\chi+q\mu L_2^2\right)\psi_I-\sigma_1k_yL_2\psi_J, 
 \end{align}
 with $m_k=\left(\Phi(r_0)\mu-k_x\right)$ and usual scaling  given by $\chi=\kappa\frac{\omega L_2^2}{(r-1)}$. Here we note the dependence of effective mass $m_k$ on the horizon value of the scalar field which in turn will be dependent upon the black hole temperature.

In the low frequency  limit $\chi\rightarrow 0$, equation (\ref{ads2eqn}) becomes
\begin{align}
\chi\partial_{\chi}\psi_I&=L_2\left[\sigma_3m+(-1)^I\sigma_1 m_k\right]\psi_I-i\sigma_2 q e_d\psi_I-\sigma_1k_yL_2\psi_J, 
\end{align}
where, $L_2 $ is the $AdS_2$ radius given by $L_2$=$\frac{1}{\sqrt{6}}$ and $\mu L_2^2=e_d$. Further we can write the above equation in a matrix form as
\begin{align}
 \chi\partial_{\chi}\psi=U\psi.
\end{align}
Here $U$ is a real constant matrix, with $\psi$ written as $(\psi_1,\psi_2)^T$. The exact form of the matrix  is given by

\begin{align*}
U\,=\,\left(\begin{array}{cccc}
 mL_2 & -m_kL_2-qe_d & 0 & -k_yL_2 \\
 -m_kL_2+qe_d  & -mL_2  & -k_yL_2 & 0 \\
0 & -k_yL_2 & mL_2 & m_kL_2-qe_d \\
-k_yL_2 & 0 & m_kL_2+qe_d & -mL_2
\end{array}
 \right).
\end{align*}

  Also $U$ is a  matrix with four eigenvalues $\mp \lambda^{I}$ ($I$=1,2). Note that there are two eigenvalues with negative sign for first $I$ component and positive sign for second component with $\lambda\,=\, \sqrt{m^2L_2^2+m_k^2L_2^2+k_y^2L_2^2-q^2e_d^2}$\,. The presence of non zero $k_y$ in this case changes the dimension of the IR CFT operator.

When block diagonalise equation (\ref{ads2eqn}) we can write as
\begin{align}
  \left[\chi\partial_{\chi}+i\sigma_2\left(\chi+qe_d\right)-(-1)^Iu_k \sigma_1\right]\psi_{1,2}=0
\end{align}
with $u_k=L_2\sqrt{m^2+(\Phi_0\mu-k_x)^2+k_y^2}$ . Since in equation (\ref{ads2eqn}) we have a mixing of four spinors, therefore by following the well known methods in  \cite{Guarrera:2011my,Faulkner:2009wj} 
 the retarded $AdS_2$ Green's function is expressed as 
 \begin{align}
 \mathcal{G}_R(\omega,k)=&e^{-i\pi\nu_k}\frac{\Gamma(-2\nu_k)\Gamma(1+\nu_k-iqe_d)}{\Gamma(2\nu_k)\Gamma(1-\nu_k-iqe_d)}\nonumber\\
 &\times\frac{(m-iu_k)L_2-\nu_k-iqe_d}{(m-iu_k)L_2+\nu_k-iqe_d}(2\omega)^{2\nu_k} .
 \end{align}\par
 where, \begin{align*}
 \nu_k= \sqrt{m^2L_2^2+m_k^2L_2^2+k_y^2L_2^2-q^2e_d^2}
 \end{align*}
 $\nu_k$ plays the role of conformal dimension of dual infrared conformal field theory operator. One can generalise for the finite temperature AdS$_2$ Green's function as in appendix \ref{appb}.  But the operator dimension $\nu_k$ would still have the same form . Further it is important see that the dimension is dependent on the temperature through the scalar field condensation, which is related to the scalar field horizon value $\Phi(r_0)$. To construct the $AdS_4$ Green's function one needs to match the solutions in $AdS_2$ usually called inner region and $AdS_4$ called outer region at their common boundary.

\section{Details calculations for model-B}\label{appa}

With the Fermions action given in (\ref{model1})
\begin{equation}\label{diracmodel1}
 \mathcal{S}^{(2)}_{Fermion}=\int d^4x\sqrt{-g}i\bar{\psi}\left(\slashed{D}-m-ip\Phi\slashed{F}\right)\psi
\end{equation}
where,\begin{align}\slashed{D}=&e^{\mu}_c\gamma^c\left(\partial_{\mu}+
\omega_{\mu}^{ab}-iq A_{\mu}\right)\nonumber\\
\slashed{F}=&\frac{1}{2}\gamma^{ab}e^{\mu}_ae^{\nu}_bF_{\mu\nu} .
 \end{align}
 The parameter $p$ is a Pauli coupling, $e^{\mu}_a ,\omega_{\mu}^{ab}$ are vielbeins and spin connection. Here, $\{a,b\}$ are tangent space indices and $\{\mu,\nu\}$ are for the bulk. Simplifying the Dirac equation from the above action with appropriate choice of gamma matrices, followed by rescaling $\psi(r,\vec{x_i})=(-gg^{r r})^{-\frac{1}{4}}e^{-i\omega t+i k.x}\tilde{\psi}(r,k)$ and rewrite $\tilde{\psi}= (\tilde{\psi_1},\tilde{\psi_2})^T$, because of symmetry we can also set $k_y=0$, then we have the following equation 
 \begin{align}\label{diraceqns0}
&r^2 \sqrt{f(r)} \partial_r \tilde{\psi}_{I} = \frac{i \sigma_2}{\sqrt{f(r)}} \left( \omega + q \mu \left(1- \frac{r_0}{r} \right) \right) \tilde{\psi}_{I}\nonumber\\& - \sigma_3  m r  \tilde{\psi}_{I} -  \sigma_1 \left( p\Phi(r) \mu \frac{r_0}{r} \pm k_x \right) \tilde{\psi}_{I}.
\end{align}
for $I\, \epsilon\,\, \{1,2\}$. We can further write $\tilde{\psi}_{I}=(\beta_I,\alpha_I)^T$ and define the ratios $\zeta_{\pm}=(\beta_{1,2}/\alpha_{1,2})$, from equation ( \ref{diraceqns0}), we get the following flow equation
\begin{align}
r^2\sqrt{f}\partial_r\zeta_{\pm}+2m\,r\zeta_{\pm} -(X_-\mp k)-(X_+\pm k)\zeta_{\pm}^2=0
\end{align}
where, 
\begin{align*}
X_{\pm}=\frac{1}{\sqrt{f}}\left(\omega+q\mu(1-\frac{1}{r})\right)\pm \frac{p\Phi\,\mu r_0}{r}
\end{align*}
Finally the Green's function is given by
\begin{align}
G_R(\omega,k)=\lim_{r\rightarrow \infty}\frac{1}{r^{2m}}\left( \begin{array}{ccc}\zeta_+ & 0  \\0 & \zeta_- \end{array}\right)
\end{align}
with in-falling boundary conditions at horizon for $\omega\neq 0$ as $\zeta_{\pm}(r=1)=i$. We will define the spectral function as 
\begin{equation}
A(\omega,k)\,=\,Tr \left[Im \left(G_R(\omega,k_x)\right)\right].
\end{equation}

\section{Finite temperature $AdS_2$ Green's function}\label{appb}

For finite temperature the AdS$_2\times S^2$ metric is given by
\begin{align*}
ds^2\,=\,\frac{L_2^2}{\zeta^2}\left(-f(\zeta)d\tau^2+\frac{d\zeta^2}{f(\zeta)}\right)+\frac{r_*^2}{L^2}d\vec{x}^2\,\,\,\,,A_{\tau}\,=\,\frac{e_d}{\zeta}\left(1-\frac{\zeta}{\zeta_0}\right)d\tau
\end{align*}
where, $f(\zeta)=1-\frac{\zeta^2}{\zeta_0^2}\,\,\,\,$. Now the Dirac equation obtained from \ref{diracmodel1} with this metric becomes

\begin{align}
\partial_{\zeta}\,\tilde{\Psi}-\frac{i\sigma_2(\omega+qA_{\tau})}{f(\zeta)}\,\tilde{\Psi}\,=\,\frac{L_2}{\zeta\,f(\zeta)}\left(m_{k\pm}\sigma_1+m\sigma_3\right)\,\tilde{\Psi}\,\,.
\end{align}
where,$m_{k\pm}\,=\,\sqrt{3}p\,\Phi(r_0)\pm\,k$\,.\par
Now this equation is exactly the equation in \cite{Faulkner:2009wj}, and solution is given by
\begin{align}
\mathcal{G}_R(\omega,T)&=\,(4\pi T)^{2\nu_k}c_{k\pm}
\end{align}
where
\begin{align}
c_{k\pm}=\frac{\Gamma(-2\nu_{k\pm})}{\Gamma(2\nu_{k\pm})}\frac{\Gamma(1+\nu_{k\pm}-iqe_d)}{\Gamma(1-\nu_{k\pm}-iqe_d)}\,\frac{\Gamma(\frac{1}{2}+\nu_{k\pm}-\frac{i\omega}{2\pi T}+iqe_d)}{\Gamma(\frac{1}{2}-\nu_{k\pm}-\frac{i\omega}{2\pi T}+iqe_d)}
 \frac{(m-im_{k\pm})L_2-iqe_d-\nu_{k\pm}}{(m-im_{k\pm})L_2-iqe_d+\nu_{k\pm}}
\end{align}
Now, the conformal dimension of the operators in the IR $=\nu_{k\pm}+\frac{1}{2}$, with $\nu_{k\pm}$ given by
\begin{align*}
\nu_{k\pm}\,=\, \sqrt{m^2L_2^2+m_{k\pm}^2L_2^2-q^2e_d^2-i\epsilon}
\end{align*}

\end{document}